\documentclass[aps,prx,twocolumn,superscriptaddress,showpacs,showkeys,
notitlepage]{revtex4-1}

\usepackage{graphicx}
\usepackage{dcolumn}
\usepackage{bm}
\usepackage{float}
\usepackage{amsmath}
\usepackage{xcolor}

\begin{document}

\title{X-ray imaging of antiferromagnetic octupole domains in Mn$_3$Sn}

%


\author{M. T. Birch}
\address{RIKEN Center for Emergent Matter Science (CEMS), Wako, Saitama 351-0198, Japan.}
\author{S. Wintz}
\address{Helmholtz-Zentrum Berlin f\"ur Materialien und Energie GmbH, Institut Nanospektroskopie, 12489 Berlin, Germany}
\author{Y. Sun}
\address{Max Planck Institute for Solid State Research
Heisenbergstrasse 1, 70569 Stuttgart, Germany}
\author{A. Kikkawa}
\address{RIKEN Center for Emergent Matter Science (CEMS), Wako, Saitama 351-0198, Japan.}
\author{M. Weigand}
\address{Helmholtz-Zentrum Berlin f\"ur Materialien und Energie GmbH, Institut Nanospektroskopie, 12489 Berlin, Germany}
\author{T. Arima}
\address{RIKEN Center for Emergent Matter Science (CEMS), Wako, Saitama 351-0198, Japan.}
\address{Department of Advanced Materials Science, University of Tokyo, Kashiwa 277-8561, Japan}
\author{Y. Tokura}
\address{RIKEN Center for Emergent Matter Science (CEMS), Wako, Saitama 351-0198, Japan.}
\address{Department of Applied Physics, The University of Tokyo, Bunkyo-ku, Tokyo 113-8656, Japan.}
\address{Tokyo College, University of Tokyo, Bunkyo-ku, Tokyo 113-8656, Japan.}

\begin{abstract}
Novel antiferromagnets with broken time reversal symmetry (TRS) have launched a new direction in spintronics research, combining the advantageous dynamical properties of conventional antiferromagnets with the controllability typically associated with ferromagnets. However, antiferromagnetic domains are notoriously challenging to image in real-space. X-ray magnetic circular dichroism (XMCD) offers a route to overcome this difficulty: XMCD contrast may be finite in TRS-breaking antiferromagnets with an appropriate magnetic space group. Here, we exploit this to image the antiferromagnetic octupole domains in a focused ion beam-fabricated device of the non-collinear antiferromagnet Mn$_3$Sn. Using scanning transmission x-ray microscopy, we spatially resolve the weak 
pre-edge XMCD contrast (of 0.2\%) that is sensitive to $T_z$, achieving a contrast resolution better than 0.02\%. We observe hysteretic switching of the octupole order through both the XMCD contrast and the corresponding anomalous Hall effect within the same device. These results confirm the bulk nature of this contrast, and establish XMCD-based microscopy as a powerful real space imaging method for TRS-breaking antiferromagnets, including altermagnets, enabling future studies of their dynamics, switching, and symmetry-tunable phenomena.
\end{abstract}

\maketitle
The search for efficient, scalable spintronic materials has driven growing interest in antiferromagnetic systems, which offer several advantages over ferromagnets: they are robust against external magnetic fields, free of stray fields, and support ultrafast spin dynamics \cite{gomonay_spintronics_2014, jungwirth_antiferromagnetic_2016, jungwirth_multiple_2018, baltz_antiferromagnetic_2018, zelezny_spin_2018, rimmler_non-collinear_2025}. However, their compensated spin structure also makes them inherently difficult to probe and control. A promising avenue to overcome this challenge is to focus on antiferromagnets that break time-reversal symmetry (TRS) -- enabling access to physical responses typically associated with ferromagnets while preserving the core benefits of antiferromagnetic order \cite{smejkal_beyond_2022, dal_din_antiferromagnetic_2024, bhowal_ferroically_2024, cheong_altermagnetism_2024}. For example, under the appropriate symmetry conditions, these TRS-breaking antiferromagnets can exhibit: the anomalous Hall effect (AHE) \cite{nakatsuji_large_2015, zhou_manipulation_2025, takagi_spontaneous_2025}, the magneto-optical Kerr effect (MOKE) \cite{higo_large_2018}, and x-ray magnetic circular dichroism (XMCD) \cite{ kimata_x-ray_2021, sakamoto_observation_2021, van_der_laan_determination_2021, amin_nanoscale_2024, yamamoto_altermagnetic_2025}, even in the absence of a net magnetization. 

Two broad classes of these unconventional antiferromagnets have emerged. The first includes non-collinear antiferromagnets such as Mn$_3$Sn, in which 120$^\circ$ spin configurations on kagome planes form a cluster magnetic octupole that breaks TRS and mirror symmetries \cite{chen_anomalous_2014, nakatsuji_large_2015, zhang_strong_2017}. The second class comprises the collinear altermagnets \cite{smejkal_emerging_2022, bai_altermagnetism_2024, tamang_newly_2024, song_altermagnets_2025, radaelli_colour_2025}, exemplified by $\alpha$-MnTe, where the two antiparallel spin sublattices are related by screw symmtery $C_6t_{1/2}$: a 60$^{\circ}$ rotation plus half translation operation. No spatial operation combined with time reversal leaves the same $\mathbf{k}$ point unchanged, so Kramers degeneracy is lifted and the collinear exchange interaction produces a spin-split band structure even though the net magnetization is (near) zero \cite{smejkal_emerging_2022,hariki_x-ray_2024}.

Finite XMCD contrast arises whenever the magnetic space group lacks a symmetry that cancels site-resolved moments \cite{yamasaki_augmented_2020}. XMCD probes the difference in absorption of left- and right-circularly polarized x-rays, and is sensitive to the projection of magnetic moments along the beam direction $z$. At the 3$d$ transition metal $L$-edges, the sum rules allow the separation of each contribution: the spin moment along the beam direction ($S_z$), the intra-atomic magnetic-dipole term that captures anisotropic spin density ($T_z$), and the orbital contribution ($L_z$), which is usually quenched and thus negligible in these metals \cite{carra_x-ray_1993}. In both altermagnetic $\alpha$-MnTe and non-collinear Mn$_3$Sn, $S_z$ is practically zero apart from minute canting-induced moments. For $\alpha$-MnTe, spin–orbit coupling pins the spins such that the $C_6t_{1/2}$ screw no longer fully cancels the $T_z$ contribution across the absorption edge \cite{yamamoto_altermagnetic_2025}. 
On the other hand, in Mn$_3$Sn, the negative vector spin chirality of the cluster octupoles yield an anisotropic spin density \cite{yamasaki_augmented_2020, kimata_x-ray_2021, sakamoto_observation_2021, sasabe_presence_2021}. Consequently, XMCD contrast in both materials is dominated by the intra-atomic magnetic-dipole term $T_z$.

These XMCD responses open a route to real-space imaging of antiferromagnetic domain structures, which is essential for understanding their symmetry-breaking nature and switching dynamics. XMCD-PEEM (photoelectron emission microscopy) \cite{amin_nanoscale_2024} and XMCD-STXM (scanning transmission x-ray microscopy) \cite{yamamoto_altermagnetic_2025} have recently been used to image altermagnetic domains in MnTe. In contrast, domain imaging in Mn$_3$Sn has so far been demonstrated via surface-sensitive techniques, including optics-based MOKE \cite{higo_large_2018, xie_magnetization_2022} and nitrogen vacancy microscopy \cite{li_nanoscale_2023}, as well as scanning thermal gradient microscopy \cite{reichlova_imaging_2019}. Here, we extend XMCD-STXM to Mn$_3$Sn, enabling direct real-space imaging of antiferromagnetic octupole domains and establishing the technique as a powerful probe of multipolar spin textures in TRS-breaking antiferromagnets.

\begin{figure}
\includegraphics[width=0.45\textwidth]{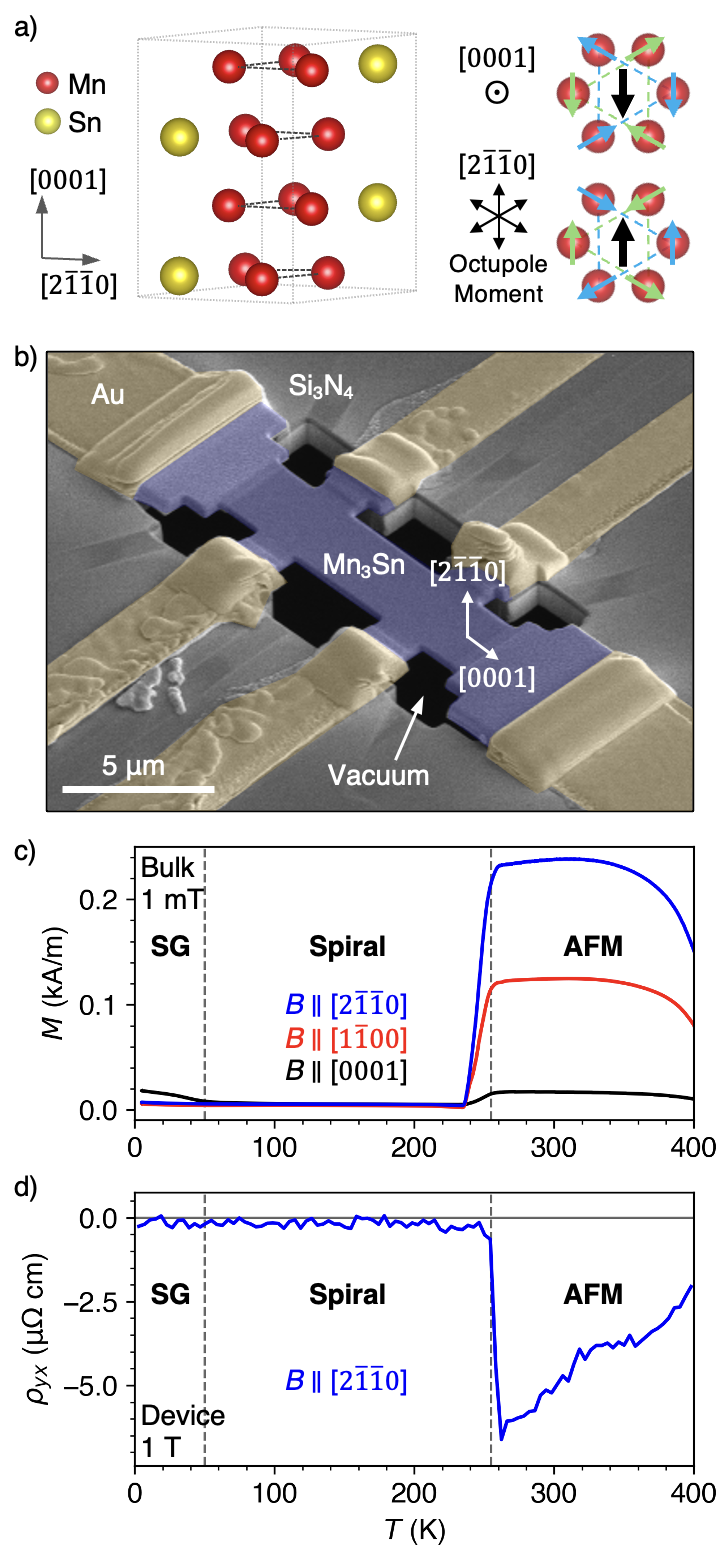}
\caption{(a) The hexagonal atomic structure and unit cell is shown. The possible orientations of the octupole moment and the corresponding arrangement of spins within each cluster of 6 Mn atoms is shown. (b) A scanning electron micrograph of the completed focused ion beam fabricated Mn$_3$Sn device. The false colours indicate the extent of the Mn$_3$Sn (blue) and the Au/Pt contact electrodes (yellow). The orientation of the crystal axes are highlighted. (c) The magnetization $M$ of the bulk Mn$_3$Sn crystal plotted as a function of temperature, measured with a $B$ field of 1 mT applied along the three high symmetry axes, as indicated by the color legend. Vertical lines indicate the transitions between the non-collinear antiferromagnetic (AFM), the antiferromagnetic spiral, and the spin glass (SG) phases. (d) The same as (c), but showing the antisymmetrised Hall resistivity $\rho_{yx}$ measured in the FIB device with an applied field of 1~T along the $[2\hat{1}\hat{1}0]$ axis.}
\label{fig_1}
\end{figure}

Mn$_3$Sn crystallizes in a hexagonal structure (space group $P6_3/mmc$), where Mn atoms form two-dimensional kagome planes stacked along the $c$-axis (Fig. 1a) \cite{nakatsuji_large_2015}. Below the N\'eel temperature of $\sim$420~K, the Mn moments adopt a coplanar 120$^\circ$ spin structure within each kagome layer, forming a non-collinear antiferromagnetic state with nearly zero net magnetization. This spin configuration gives rise to a ferroic ordering of magnetic octupoles, which serves as the primary order parameter in this phase. The octupole arises from the coherent superposition of spins on the six Mn atoms in the unit cell \cite{hayami_unified_2024}, and encodes the TRS breaking responsible for the observed anomalous Hall and dichroic effects \cite{chen_anomalous_2021}. Due to a slight spin canting, a small net magnetization enables the orientation of the octupole moment to be controlled within the hexagonal plane by an externally applied magnetic field. Mn$_3$Sn has been reported in two distinct types \cite{park_magnetic_2018, singh_higher_2024}: i) type A, where below a temperature of $\sim$260~K the coplanar antiferromagnetism develops a helical modulation along the c axis and the spontaneous net moment is lost \cite{park_magnetic_2018, chen_intertwined_2024}; ii) type B, where instead the typical AFM phase persists to low temperatures \cite{nakatsuji_large_2015}. In both cases, a spin glass state is thought to emerge below $\sim$50~K. Prior studies indicate that type A behavior typically occurs in stoichiometric crystals, while type B is associated with Mn-rich compositions \cite{park_magnetic_2018}.

We grew a single crystal of Mn$_3$Sn using a self-flux growth method, targeting type A behavior, starting from a Sn-rich composition of 2.33:1 (following the recipe in Ref. \cite{sung_magnetic_2018}). After aligning and cutting a suitable section, we utilized a focused ion beam (FIB) instrument to fabricate a Hall bar device, as shown in Fig. 1b. A thin lamella was first extracted and thinned to a thickness of $\sim$350~nm, using a standard FIB lift-out procedure. This lamella, with the $[2\bar{1}\bar{1}0]$ axis oriented out of plane, was transferred to a Si$_3$N$_4$ membrane featuring an aperture surrounded by pre-patterned Au electrodes. Electrical contact was established using the in situ Pt deposition, with the $[0001]$ axis oriented along the current direction. The Hall bar was then shaped via ion beam milling, producing the final free-standing geometry. Next, the sample was polished at a low angle with the ion beam set to a lower acceleration voltage of 5~kV, reducing the thickness to a final value of $\sim$250~nm. Finally, an Al$_2$O$_3$ capping layer was deposited via atomic layer deposition to prevent degradation of the Mn$_3$Sn, which was otherwise found to readily oxidize in ambient conditions.

The temperature-dependent magnetization $M$ of the Mn$_3$Sn single crystal was measured using a Quantum Design MPMS3, with an external magnetic field $B$ applied along the three high-symmetry directions of the hexagonal lattice (Fig. 1c). The steep decrease in $M$ at around 260 K indicates the onset of the antiferromagnetic spiral phase, confirming that our crystal is of type A. This behavior was corroborated by transport measurements of the Mn$_3$Sn device using a Quantum Design PPMS: Fig. 1d shows the Hall resistivity $\rho_{yx}$ as a function of temperature, under a 1~T magnetic field applied along the $[2\bar{1}\bar{1}0]$ axis, revealing the sharp drop of $\rho_{yx}$ at the onset of the antiferromagnetic spiral-order.

\begin{figure}
\centering
\includegraphics[width=0.45\textwidth]{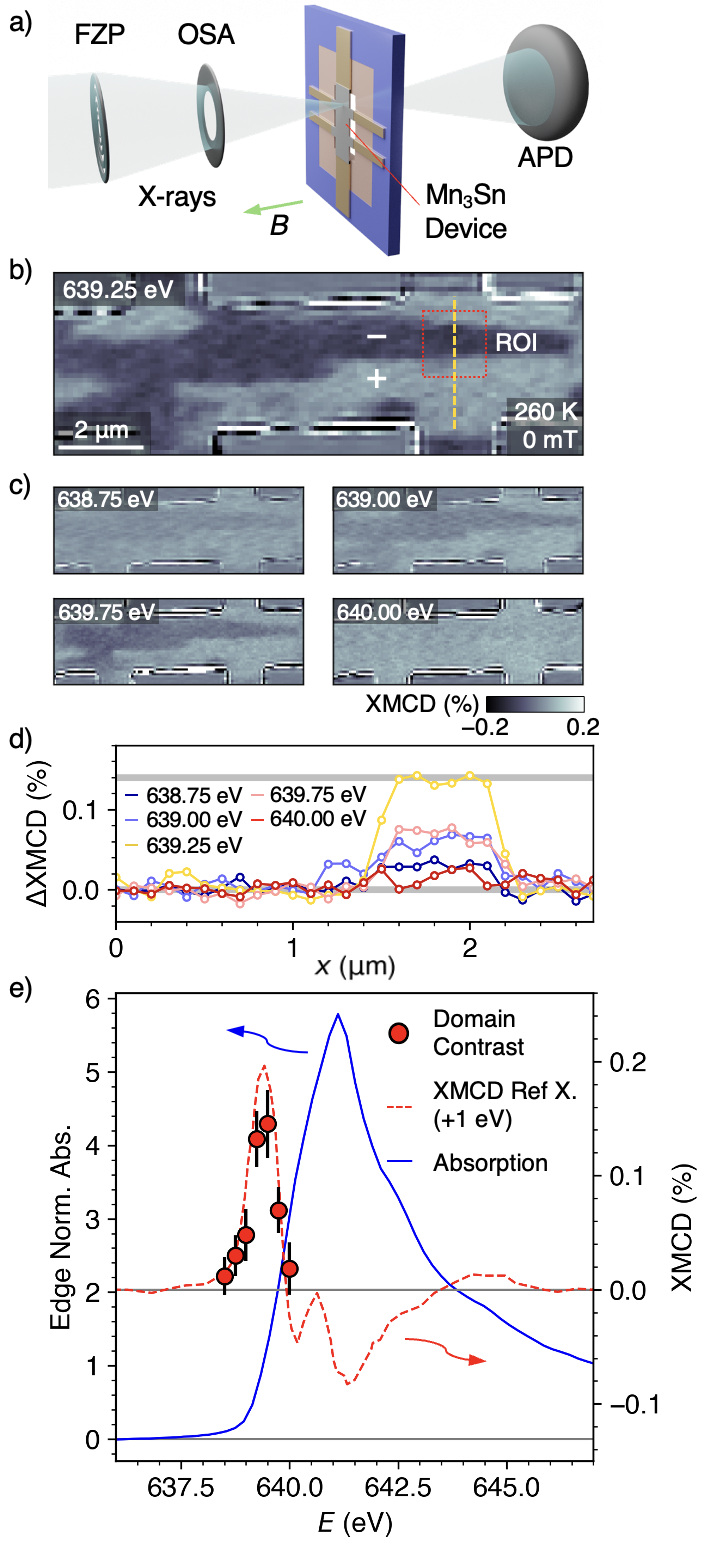}
\caption{(a) Illustration of the STXM setup, showing the Fresnel zone plate (FZP), order selecting aperture (OSA), and the avalanche photodiode (APD). (b) An XMCD contrast image of the antiferromagnetic octupole domains in the Mn$_3$Sn device at 260~K and with a nominal x-ray energy $E$ of 639.25~eV. (c) Further example XMCD contrast images, acquired at a range of $E$ around the Mn L$_3$ edge. (d) Linescans of the XMCD contrast at each $E$ over the octupole domain indicated by the orange line in the region of interest (ROI) in (b). (e) The average XMCD contrast within the ROI at each energy was extracted, and plotted as a function of $E$ (red circles), where the error is standard error. The absorption measured in a thinner Mn$_3$Sn sample, showing the Mn L$_3$ edge (blue line). Our observed XMCD contrast is in strong agreement with that measured in reference \cite{sakamoto_observation_2021} (red-dashed line).}
\label{fig_2}
\end{figure}

STXM measurements were performed using the MAXYMUS instrument at BESSY II, equipped with a liquid helium cryostat for temperature control. The imaging setup is illustrated in Fig. 2: the soft x-ray beam was focused to a $\sim$25~nm spot size using a Fresnel zone plate, and the first diffraction order light was selected by an order selecting aperture. The sample was raster-scanned through the focal spot pixel by pixel, and the transmitted beam intensity was measured by an avalanche photodiode. The Mn$_3$Sn device was mounted within the vacuum chamber, and an external magnetic field $B$ was applied along the out-of-plane direction by the arrangement of four permanent magnets. 

To observe the octupolar XMCD contrast, the energy of the beam was set to a nominal value of 639.25~eV, just below the peak of the Mn L$_3$ edge. The sample was first cooled to 200~K, entering the spiral AFM phase where the ferroic octupole order is suppressed, and then warmed to 260~K to re-enter the non-collinear AFM phase. This thermal cycling reliably initialized a multidomain state. The resulting octupole domain configuration is shown in Fig. 2b, where bright and dark contrast corresponds to the spin clusters oriented parallel or antiparallel to the x-ray beam. The XMCD contrast was obtained by imaging the sample with left- and right-circularly polarized light at a 100~nm pixel resolution. Each image was normalized by the intensity of the straight-through beam. The logarithm of each image was then computed, and then the resulting images were subtracted from one another to yield the XMCD contrast. During our measurements we observed only two contrast levels in the domain structure. While in principle the octupole can be oriented along any of the $[2\bar{1}\bar{1}0]$ axes, it is possible that residual strain in the device selects one of these axes \cite{higo_perpendicular_2022}.

\begin{figure*}
\centering
\includegraphics[width=0.75\textwidth]{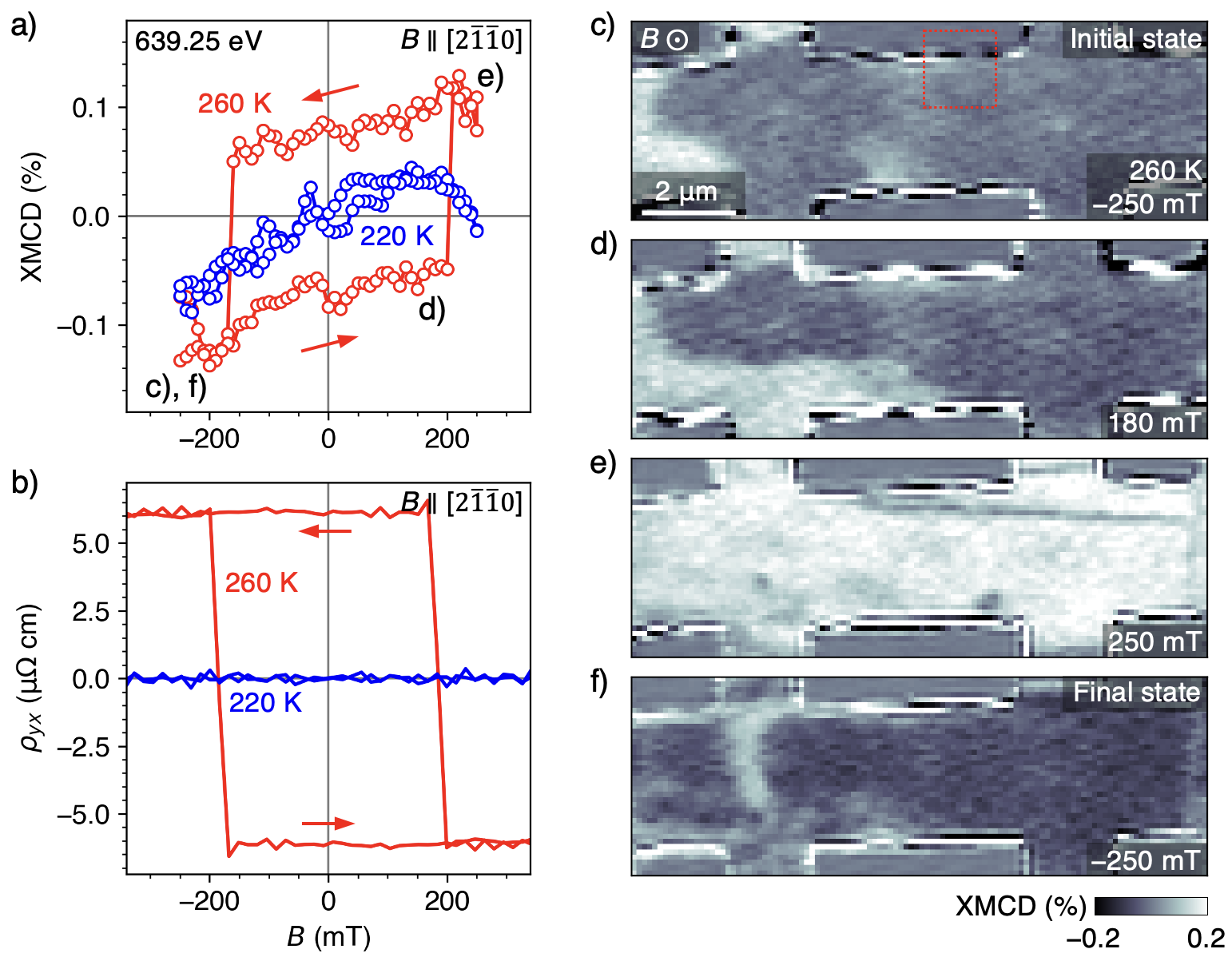}
\caption{(a) The XMCD contrast acquired by averaging over the region of interest of the Mn$_3$Sn device (highlighted in Fig. 2(b)), plotted as a function of the applied magnetic field $B$, at 260~K (red) and 220~K (blue). (b) The Hall resistivity $\rho_yx$ acquired as a function of $B$, measured in the Mn$_3$Sn device, at 260~K (red) and 220~K (blue). (c)-(f) Sequential XMCD contrast images of the octupole domain state within the device at a selection of applied $B$, as indicated. The red box indicates the area probed in (a).}
\label{fig_3}
\end{figure*}

To confirm that the observed XMCD contrast originates from the octupole order via the $T_z$ term, we imaged the same domain configuration at several photon energies around the Mn L$_3$ pre-edge region, as shown in Fig. 2c. To quantify the XMCD contrast sensitivity, we extracted line scans across the adjacent domains (shown as the orange dashed line in Fig. 2b), with results shown in Fig.~2d. These line profiles show clear contrast reversals, with a maximum contrast of approximately 0.2\% at the ideal energy of 639.25~eV. The signal exhibits a noise floor below 0.02\%, enabling robust identification of weak octupole domain contrast. This high sensitivity was achieved using a dwell time of 2~ms per pixel, reflecting the imaging performance of the MAXYMUS STXM instrument at BESSY II.

At each photon energy, we averaged the XMCD contrast between the two domains within the region of interest marked by the red box in Fig. 2b. These values are plotted as the red circles in Fig. 2d. The primary device was too thick for the entire spectrum to be measured, so the blue line shows the x-ray absorption spectrum of a thinner Mn$_3$Sn sample measured across the entire Mn L$_3$ edge. A key signature of the octupole domain contrast is the positive sign of the XMCD signal \cite{kimata_x-ray_2021, sakamoto_observation_2021} -- ferromagnetic Mn would typically show negative XMCD contrast at the L$_3$ edge. To highlight this positive signal, we overlay a reference XMCD spectrum for Mn$_3$Sn from Ref. \cite{sakamoto_observation_2021}, plotted as the red dashed line. Because the two datasets were acquired at different synchrotrons, we applied an energy shift of ~+1~eV to the data from Ref. \cite{sakamoto_observation_2021} to match the energy calibration, determined by aligning the XMCD peak position. The quantitative agreement is striking: both our domain contrast data and the reference spectrum exhibit a positive XMCD signal with a maximum value of $\sim$0.2\% at 639.25~eV. This confirms that the observed contrast originates from the time-reversal symmetry breaking of the spin-anisotropic octupole order.

The maximum available magnetic field available in our setup ($\pm$250 mT) is sufficient to control the octupole domain state within the Mn$_3$Sn device. Fig. 3a shows the XMCD contrast in the region of interest as a function of $B$. The XMCD contrast was acquired using the same method as described for the images above. At 260~K, the contrast exhibits a clear hysteresis loop associated with octupole domain switching. In contrast, at 220~K -- where the system is in the antiferromagnetic spiral phase -- only a weak linear response is observed. The corresponding Hall effect data is shown in Fig. 3b, and closely tracks the XMCD signal. The possibility for both a finite XMCD and AHE in TRS-breaking antiferromagnets is given by the same symmetry rules \cite{hariki_x-ray_2024, liu_different_2025, radaelli_tensorial_2024, bhowal_ferroically_2024}. However, while the XMCD considers the orientation of spins (or anisotropic spin density) relative to the beam axis, the AHE considers the mapping of the Berry curvature $\Omega(\mathbf k)\!\rightarrow\!-\Omega(\mathbf k)$ throughout the Brillouin zone. Thus, although allowed by the same symmetry, their magnitudes are not coupled \cite{hariki_x-ray_2024}: the XMCD is governed mainly by on-site matrix elements such as the intra-atomic magnetic-dipole term $T_z$, whereas the AHE depends on the momentum-space distribution of $\Omega(\mathbf k)$. 

Representative images acquired during the magnetic field sweep are shown in Fig. 3c–f, revealing clear reversal of the octupolar domain contrast at each field extreme. The switching is not fully complete -- likely because the maximum available field in the x-ray microscope is insufficient to fully saturate the antiferromagnetic octupole domain state, although we cannot rule out imperfections from the FIB milling and Pt deposition processes. This incomplete saturation may also explain the slight different in the switching field for the up and down field sweep directions. While the majority of the sample switches sharply, a mixed-domain configuration is often observed during the sweep, with octupolar domain walls clearly visible.

We have demonstrated real-space imaging of antiferromagnetic octupole domains in a focused-ion beam fabricated Mn$_3$Sn device, and directly correlated the domain contrast with the anomalous Hall effect. This was made possible by the high XMCD contrast sensitivity we achieved: better than 0.02\%. Our results extend the capabilities of XMCD contrast imaging to the study of multipolar order in TRS-breaking antiferromagnets. Compared to other imaging techniques, XMCD-STXM offers a unique combination of element specificity, depth sensitivity, and compatibility with buried or encapsulated devices. Optical methods such as MOKE are limited to surface sensitivity and require finite Kerr rotation, while electron microscopy techniques generally lacks magnetic contrast in antiferromagnets unless combined with advanced instrumentation and precise crystallographic alignment \cite{kohno_real-space_2022}. In contrast, XMCD-STXM directly accesses the symmetry-breaking magnetic multipole order within the bulk, and can be extended to time-resolved pump–probe studies with picosecond temporal resolution.

\begin{acknowledgments}
The authors thank Helmholtz-Zentrum Berlin for the allocation of synchrotron radiation beamtime at the BESSY II synchrotron. We thank the CEMS Semiconductor Science Research Support Team for the use of cleanroom facilities. We are grateful to X. Z. Yu and her team for the use of the Helios 5UX focused ion beam system. We thank Y. Taguchi for helpful discussions. MTB acknowledges the RIKEN Special Postdoctoral Research Fellowship scheme, and the RIKEN Incentive Research Project scheme.
\\
\end{acknowledgments}

\bibliography{sample}

\end{document}